\documentclass[a4paper,11pt]{article}
\usepackage{pos}
\usepackage{graphicx}
\usepackage{wrapfig}
\let\oldbibliography\thebibliography
\renewcommand{\thebibliography}[1]{%
  \oldbibliography{#1}%
  \setlength{\itemsep}{-4pt}%
}
\newcommand{\sqrts}{\sqrt{s_\textrm{NN}}}
\newcommand{\kompost}{K\o{}MP\o{}ST}

\title{Event-plane decorrelation of photons produced in the early stage of heavy-ion collisions}
\ShortTitle{Event-plane decorrelation of early photons in heavy-ion collisions}

\author[a]{Charles Gale}
\author*[b]{Jean-Fran\c{c}ois Paquet}
\author[c]{Bj\"orn Schenke}
\author[d]{Chun Shen}
\affiliation[a]{Department of Physics, 3600 University Street, Montr\'eal, QC, H3A 2T8, Canada}
\affiliation[b]{Department of Physics, Duke University, Durham, NC 27708, USA}
\affiliation[c] {Physics Department, Brookhaven National Laboratory, Upton, NY 11973, USA}
\affiliation[d]{Department of Physics and Astronomy, Wayne State University, Detroit, MI 48201, USA \\
RIKEN BNL Research Center, Brookhaven National Laboratory, Upton, NY 11973, USA}

\emailAdd{gale@physics.mcgill.ca}
\emailAdd{jeanfrancois.paquet@duke.edu}
\emailAdd{bschenke@quark.phy.bnl.gov}
\emailAdd{chunshen@wayne.edu}

\abstract{We study photon production in the early stage of heavy-ion collisions using a multistage model combining IP-Glasma, \kompost{} and relativistic hydrodynamics. 
We discuss the small momentum anisotropy of these photons, highlighting the role of the photon-hadron event-plane decorrelation.
We comment that this singular characteristic of early photons could be used to provide dynamical information on the complex pre-hydrodynamics phase of heavy-ion collisions.}

\FullConference{%
  HardProbes2020\\
  1-6 June 2020\\
  Austin, Texas
}

\begin{document}
\maketitle

\paragraph{Introduction}

\begin{figure}[bh]
	\centering
	\includegraphics[width=0.99\linewidth]{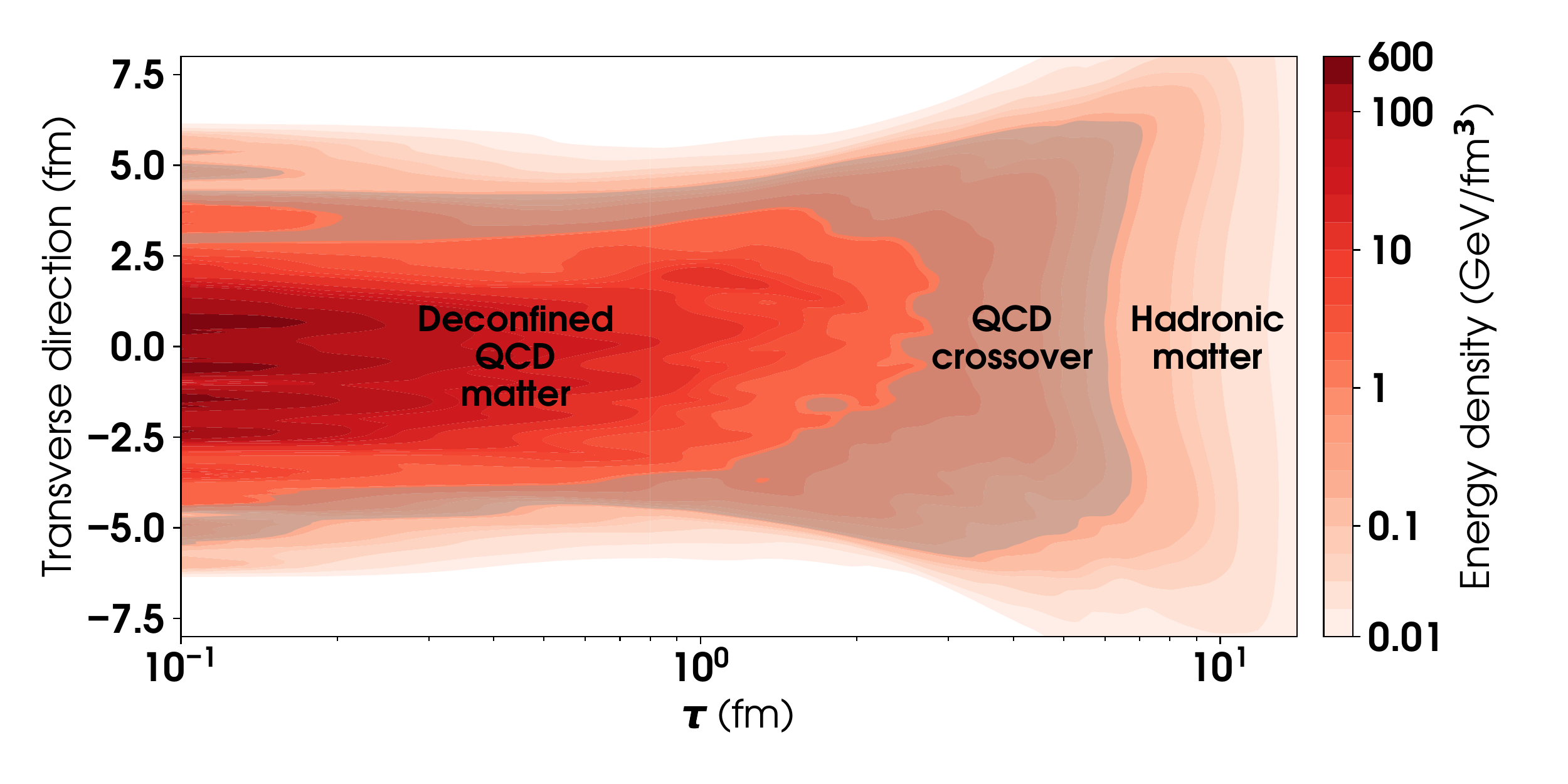}
	\vspace{-0.2cm}
	\caption{Illustration of the energy density in the transverse plane, as a function of time. This example is for a single Au-Au collision at $\sqrts=200$~GeV. The shaded grey area labeled ``QCD crossover'' corresponds to energy densities between 0.234 and 1.77 GeV/fm$^3$ (temperatures 150--200~MeV for an equilibrium QCD plasma).}
	\label{fig:energy_density_profile}
\end{figure}

Relativistic collisions of heavy nuclei produce a short-lived medium of deconfined nuclear matter. Shortly after the impact of the nuclei, the local energy density of this deconfined medium can reach hundreds of GeV/fm$^3$.
An illustration of such an energy density profile is shown in Fig.~\ref{fig:energy_density_profile}, as a function of time and a direction transverse to the collision axis.
At high densities, the deconfined medium takes the form of a complex excited state of strongly-interacting coloured degrees of freedom. 
This matter expands rapidly in the vacuum of the collider's beam pipe, passing through the QCD crossover as it becomes more dilute and finally reconfining into hadrons.

The evolution of the medium
produced in heavy-ion collisions
is governed by the strong nuclear interaction.
Yet a large number of degrees of freedom produced in the collisions carry an electric charge, leading to continuous photon emission throughout the collision.
In this work, we discuss photons emitted in the early stages of the plasma. We focus on a distinguishing feature of their momentum distribution: the decorrelation of their event plane with that of soft hadrons.

\paragraph{Spacetime description of heavy-ion collisions}

It is likely that the deconfined and crossover phase of the collision --- the darker shades in Fig.~\ref{fig:energy_density_profile} ($\tau \lesssim 5-7$~fm) --- do not admit a general quasi-particle picture, a consequence of the strongly-interacting nature of the medium.
More commonly, the deconfined medium is described in terms of macroscopic quantities, such as the energy density (depicted in Fig.~\ref{fig:energy_density_profile}) and the flow velocity profile of the medium. 
For times larger than $\mathcal{O}(1$~fm), there is good evidence that the spacetime evolution of the energy density and fluid velocity can be described with relativistic hydrodynamics~\cite{Gale:2013da}.
This hydrodynamics description cannot be extended to arbitrarily early times, and the ``pre-hydrodynamics'' stage of heavy-ion collisions contains rich physics that must be modelled independently from hydrodynamics. 

In this work, we model the ``pre-hydrodynamics'' stage with IP-Glasma~\cite{Schenke:2012wb,Schenke:2012hg} and \kompost{}~\cite{Kurkela:2018wud,Kurkela:2018vqr}. We first use IP-Glasma to describe the incoming nuclei using the Color-Glass-Condensate effective theory and subsequently evolve the deconfined matter for up to $\tau=0.1$~fm by solving the Yang-Mills equations.
The energy-momentum tensor of IP-Glasma at $\tau=0.1$~fm is then used to initialize the \kompost{} framework, which evolves it up to $\tau=0.8$~fm. 
In \kompost{}, the energy-momentum tensor is divided into a uniform background --- defined locally from the causal circle around this point --- and linearised fluctuations atop this background. The evolution of this background follows a simple scaling relation~\cite{Kurkela:2018wud,Kurkela:2018vqr,Giacalone:2019ldn}.
Fluctuations are propagated with response functions that have been calculated in QCD kinetic theory~\cite{Kurkela:2018wud,Kurkela:2018vqr}.
At $\tau=0.8$~fm,  the energy-momentum tensor from \kompost{}  is used to initialize the relativistic viscous hydrodynamic simulation, which includes both shear and bulk viscosities, as described in Ref.~\cite{Gale:2020xlg}. The energy density profile from Fig.~\ref{fig:energy_density_profile} is the result of the \kompost{} and hydrodynamic evolution.

\paragraph{Evaluating photon emission}

Photon emission at lower energy densities --- the lighter shades in Fig.~\ref{fig:energy_density_profile} --- is relatively well understood, since the medium takes the form of an interacting gas of hadrons.
A number of hadronic channels lead to the production of photons, many of which have been calculated using hadronic effective theory (e.g. Ref.~\cite{Turbide:2003si}). 
A source of uncertainty originates from the exact mapping between the energy-momentum tensor of hydrodynamics and the hadronic momentum distribution, an uncertainty that propagates to the calculation of photon emission in the form of ``viscous corrections'' to the thermal photon emission rates.
In this work, we use the same hadronic photon emission rates and the same viscous correction as in Ref.~\cite{Paquet:2015lta}.

Photon emission at higher energy densities is more challenging to calculate because of the strongly-interacting nature of the deconfined medium.
Intuition originates in part from our understanding of weakly-coupled deconfined nuclear matter. In a weakly-coupled quark-gluon plasma, photons are produced through elementary processes such as Compton scattering ($q g \to q \gamma$) and quark-antiquark annihilation ($q \bar{q} \to g \gamma$), along with more complex channels through bremsstrahlung and inelastic pair annihilation~\cite{Arnold:2001ms}.
These channels can be used to calculate photon emission rates for deconfined nuclear plasma.
Strictly speaking, these rates are only valid for energy densities asymptotically higher than those encountered in heavy-ion collisions, where the strong coupling constant $g_s$ is small.
Moreover these rates are only known for plasmas relatively close to local thermal equilibrium.
The challenge presented by the higher density regions of Fig.~\ref{fig:energy_density_profile} is twofold: (i) the higher density regions, while more weakly coupled, also tend to deviate more significantly from thermal equilibrium; while (ii) the intermediate density regions are thought to be closer to equilibrium, but are also more strongly coupled. In this work, we follow previous studies such as Ref.~\cite{Paquet:2015lta}: we calculate photon production in the deconfined phase with photon emission rates calculated in the weakly-coupled limit and then extrapolated to $g_s=2$. Corrections due to deviations from thermal equilibrium are also treated as in Ref~\cite{Paquet:2015lta}, for both shear and bulk viscosity. We use the exact same approach to compute photons in the \kompost{} and hydrodynamic phases. Other sources of photons, such as prompt photons, are left out at the moment.

\paragraph{Early photon emission and photon-hadron correlations}

\begin{figure}[tbh]
	\centering
	\includegraphics[width=0.49\linewidth]{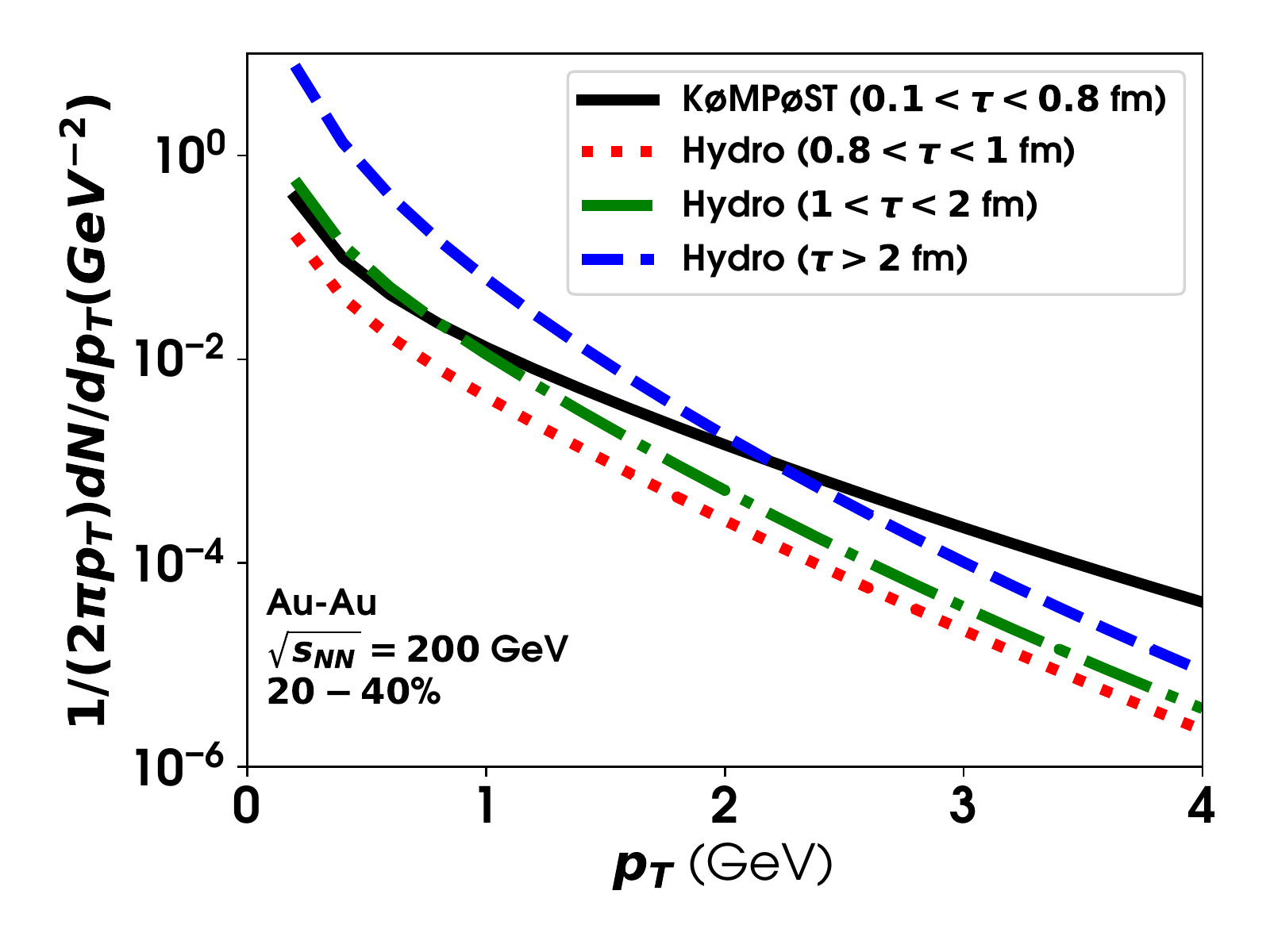} \llap{\parbox[b]{0.4in}{(a)\\\rule{0ex}{1.34in}}}
	\hfill
	\includegraphics[width=0.49\linewidth]{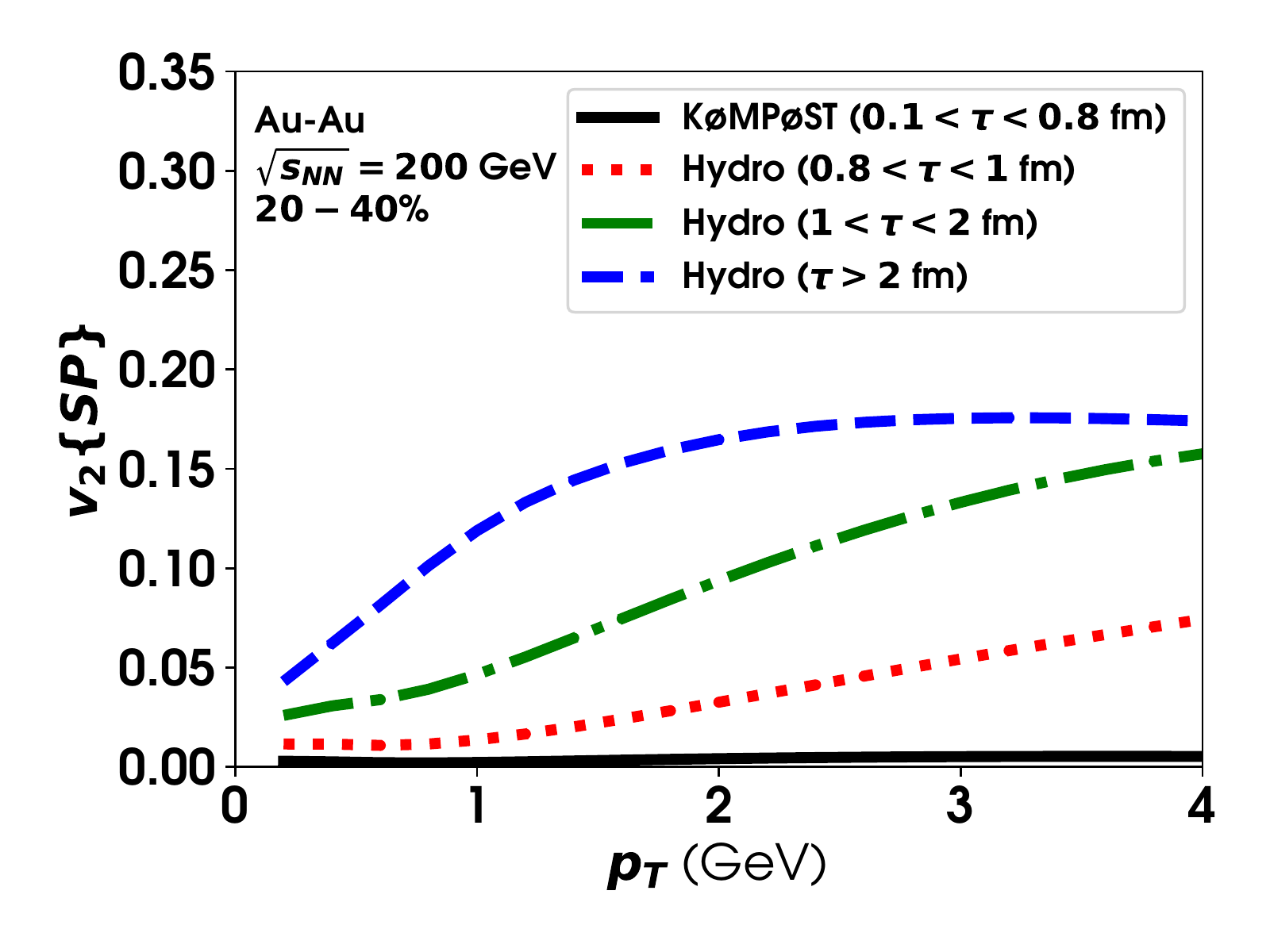}
	\llap{\parbox[b]{0.4in}{(b)\\\rule{0ex}{1.27in}}}
	\caption{(a) Spectra and (b) $v_2\{SP\}$ (Eq.~\ref{eq:v2SP}) of photons produced at different time intervals.}
	\label{fig:spectra_v2_photons}
\end{figure}

The spectra of photons produced at different time intervals in \kompost{} and the hydrodynamics is shown in Fig.~\ref{fig:spectra_v2_photons}(a). The corresponding photon momentum anisotropy is shown in Fig.~\ref{fig:spectra_v2_photons}(b). As a general feature, photons produced in the earlier stage of the collision contribute predominantly at higher $p_T$. They also have a small momentum anisotropy. Recall that the momentum anisotropy of hadrons originates from the development of an anisotropic flow velocity, itself a consequence of an asymmetry in the spatial energy deposition after the nuclei's impact. This flow velocity anisotropy takes time to build; we see on Fig.~\ref{fig:spectra_v2_photons}(b) that photons emitted earlier in the collision do have a smaller momentum anisotropy.

\begin{wrapfigure}{l}{0.5\textwidth}
	\centering
	\vspace{-0.4cm}
	\includegraphics[width=\linewidth]{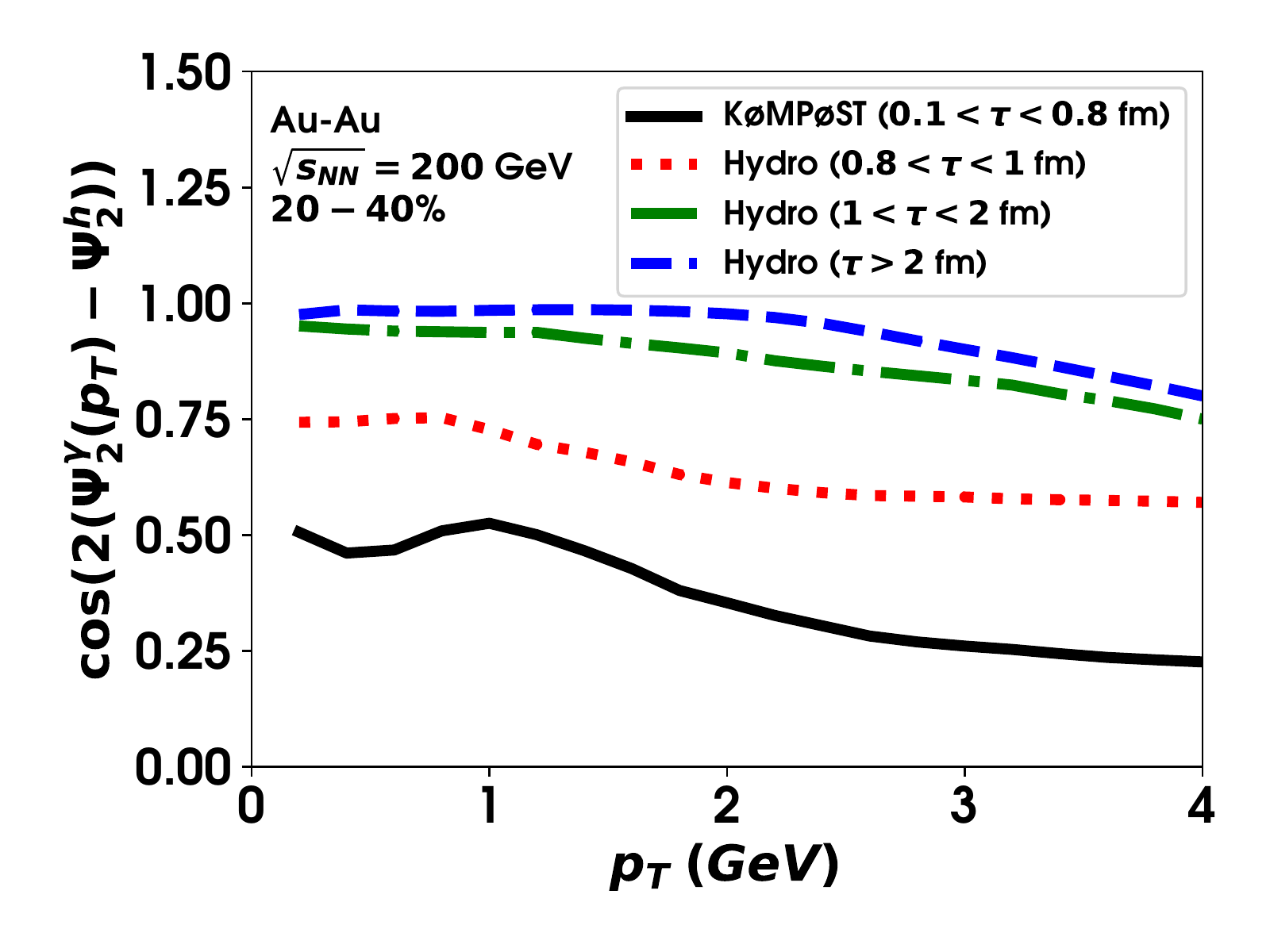} 
	\caption{Correlation between the hadron and photon event plane, as quantified by $\cos(2(\Psi_2^{\gamma}-\Psi_2^{h}))$.}
	\label{fig:cos_psi}
\end{wrapfigure}

\noindent The anisotropy in the flow velocity during the pre-hydrodynamics phase is considerably smaller than that  developed in the hydrodynamics phase.
Moreover the angular distribution of the flow velocity in the IP-Glasma phase is effectively uncorrelated with that developed in the hydrodynamic phase. 
This can be quantified by comparing the event plane angle of soft hadrons with that of photons produced at different stages of the collision. Recall that the measured photon anisotropy is a photon-hadron correlation given by
\begin{equation}
v_n\{SP\}(p_T^\gamma)=\frac{\langle v_n^{\gamma}(p_T^\gamma) v_n^{h} \cos(n(\Psi_n^{\gamma}(p_T^\gamma)-\Psi_n^{h})) \rangle}{\sqrt{\langle ( v_n^{h} )^2 \rangle}}
\end{equation}
with
\begin{equation}
v_n(p_T) e^{i n \Psi_n(p_T)} =\left [\int  dy d\phi \left( p^0 \frac{d^3 N}{d^3 p} \right) e^{i n\phi} \right] \Big{/} \left[\int dy  d\phi \left( p^0 \frac{d^3 N}{d^3 p} \right) \right] \\
\label{eq:v2SP}
\end{equation}
where $p_T$ is the transverse momentum, $y$ the momentum rapidity and $\phi$ the transverse azimuthal angle, and  $p^0 d^3 N^{h/\gamma}/d^3 p$ is the momentum distribution of photons or hadrons for a single collision event. 
The hadronic $v_n^{h}$ and $\Psi_n^{h}$ are integrated over $p_T$. 
The event plane $\Psi^{h/\gamma}_n$ of photons or hadrons enters in $v_n\{SP\}$ through $\cos(n(\Psi_n^{\gamma}-\Psi_n^{h}))$. This factor is plotted in Fig.~\ref{fig:cos_psi} for photons emitted at different times in the medium. At higher $p_T$, where photons emitted at early times tend to dominate, the value of $\cos(2(\Psi_2^{\gamma}-\Psi_2^{h}))$ is around $0.25$. This decorrelation of the photon and hadron event planes, combined with the overall small $v_2$ of these early stage photons, results in the very small $v_2\{SP\}$ seen in Fig.~\ref{fig:spectra_v2_photons}(b).

\paragraph{Summary}

Evaluating photon emission from the early stages of heavy-ion collisions is currently a developing field of inquiry. While uncertainties are still significant, there is accumulating evidence that photons produced at the earlier stage of heavy-ion collisions 
have a small $v_2$ and an event plane that is decorrelated with that of soft hadrons. 
Unless a mechanism can be found to produce simultaneously a larger $v_2$ and a stronger correlation with soft hadrons, these photons are unlikely to provide a solution to the ``direct photon puzzle''. Yet these same characteristics can be an asset, distinguishing them from the larger number of photons produced in the later stages of the collisions, and providing an additional probe of the complex early stage of heavy-ion collisions.

\paragraph{Acknowledgments}

This work was supported by the U.S. Department of Energy (DOE) under grant numbers DE-FG02-05ER41367, DE-SC0012704, and DE-SC0013460, by National Science Foundation (NSF) under grant number PHY-2012922, and by the Natural Sciences and Engineering Research Council of Canada. This research used resources of the National Energy Research Scientific Computing Center (NERSC), a U.S. DOE Office of Science User Facility operated under grant number DE-AC02-05CH11231.

%
%
%
%
%
\bibliographystyle{JHEP}
\bibliography{biblio}

\end{document}